\documentclass[fleqn,10pt]{elsarticle}
\usepackage{lineno}
\usepackage{multirow}
\usepackage{subfigure}
\usepackage{graphicx}
\usepackage{graphics}
\usepackage{epstopdf}
\usepackage{extarrows}
\usepackage{xcolor}
\usepackage{geometry}
\usepackage{url}
\usepackage{float}

\usepackage{mathrsfs}
\usepackage{subfigure}
\usepackage{amssymb}
\usepackage{amsmath}
\usepackage{color}
\usepackage{ulem}
\usepackage{array}
\usepackage{longtable}
\usepackage{lscape}
\usepackage{multicol}
\numberwithin{equation}{section}
\allowdisplaybreaks

\modulolinenumbers[5]

\begin{document}

\renewcommand{\vec}[1]{\boldsymbol{#1}}

\begin{frontmatter}

\title{Modelling infectious disease transmission dynamics in conference environments: An individual-based approach}

\author[add1,add5,add7]{Xue Liu}
\author[add2,add7]{Yue Deng}
\author[add3,add6]{Jingying Huang}
\author[add2]{Yuhong Zhang}
\author[add1,add4]{Jinzhi Lei\corref{correspondingauthor}}
\address[add1]{School of Mathematical Sciences, Tiangong University, Tianjin, 300387, China. }
\address[add2]{School of Software, Tiangong University, Tianjin, 300387, China.}
\address[add3]{School of Computer Science and Technology, Tiangong University, Tianjin, 300387, China} 
\address[add4]{Center for Applied Mathematics, Tiangong University, Tianjin, 300387, China.} 
\address[add5]{X. Liu is now at the School of Mathematics, Shandong University.}%
\address[add6]{J. Huang is now at Shanghai Qinhe Web Technology Software Development Co., Ltd. }
\address[add7]{These authors contributed equally}
\cortext[correspondingauthor]{Corresponding author: jzlei@tiangong.edu.cn (J. Lei)}


\begin{abstract}
The global public health landscape is perpetually challenged by the looming threat of infectious diseases. Central to addressing this concern is the imperative to prevent and manage disease transmission during pandemics, particularly in unique settings. This study addresses the transmission dynamics of infectious diseases within conference venues, presenting a computational model designed to simulate transmission processes within a condensed timeframe (one day), beginning with sporadic cases. Our model intricately captures the activities of individual attendees within the conference venue, encompassing meetings, rest intervals, and meal breaks. While meetings entail proximity seating, rest and lunch periods allow attendees to interact with diverse individuals. Moreover, the restroom environment poses an additional avenue for potential infection transmission. Employing an individual-based model, we meticulously replicated the transmission dynamics of infectious diseases, with a specific emphasis on close-contact interactions between infected and susceptible individuals. Through comprehensive analysis of model simulations, we elucidated the intricacies of disease transmission dynamics within conference settings and assessed the efficacy of control strategies to curb disease dissemination. Ultimately, our study proffers a numerical framework for assessing the risk of infectious disease transmission during short-duration conferences, furnishing conference organizers with valuable insights to inform the implementation of targeted prevention and control measures.
\end{abstract}


\end{frontmatter}

\section{Introduction}

The profound impact of the COVID-19 pandemic on global public health and economic stability over the past few years underscores the critical need for comprehensive reflection on preventive measures and disease control strategies \cite{WorldHealthOrganization2020a, WorldHealthOrganization2020b}. As the pandemic era wanes, we must scrutinize and enhance our understanding of measures to curb the spread of infectious diseases. Despite the extensive body of literature on epidemic dynamics and forecasting, there remains a paucity of research on disease transmission dynamics within specific contexts. Notably, evaluating the potential for disease outbreaks following large-scale international events such as conferences and business meetings is paramount in an era characterized by interconnectedness and globalization.

Numerous mathematical models have been proposed to predict the evolution of epidemic dynamics and evaluate the efficacy of various prevention and control strategies. These models often take the form of differential equations based on the SIR (susceptible-infectious-recovered) or SEIR (susceptible-exposed-infectious-recovered) frameworks \cite{2020Epidemic, 2020Estimation, 2020A, 2020Modeling,Deng:2021gg}. Additionally, data-driven models have emerged to forecast dynamics through analysis of reported data \cite{2021Effectiveness, 2020DataDriven, 2017Using, 2021From, 2020Data}. While compartmental and data-driven models are valuable for examining community transmission during epidemics, they may not adequately capture sporadic transmission within specific settings. To address the stochastic dynamics of disease transmission arising from close interpersonal contact, researchers often employ individual-based models, which simulate the epidemic progression for each individual \cite{Ajelli2010ComparingLC,Xu2021EffectivenessON, Chang2020ModellingTA,Wu:2023kg}.

The present study endeavors to develop a computational model capable of simulating the spread of infectious diseases within the confines of a one-day conference venue, with a specific focus on respiratory diseases triggered by viral infection. Our model incorporates various activities--such as meetings, breaks, and meal times--that participants typically engage in, facilitating dynamic contact patterns as individuals interact with diverse attendees throughout the day. Moreover, we acknowledge the potential for infection transmission in restroom facilities due to environmental factors. The individual-based model proposed in this study offers a quantitative approach to assessing the risk of disease transmission during short-duration conferences and evaluating diverse prevention and control strategies to mitigate infection spread.

\section{Methods}

\subsection{Individual-based model}

\begin{figure}[htbp]
  \centering
{\includegraphics[width=13cm]{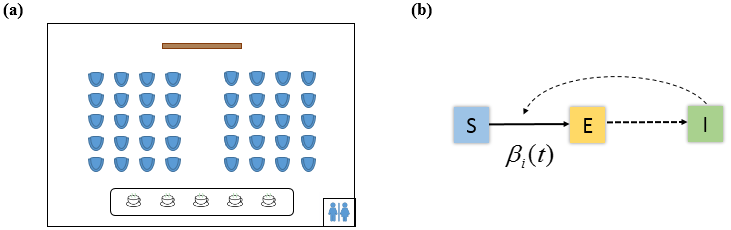}}
  \caption{\textbf{(a)} Status map of the conferene venue. \textbf{(b)} Schematic representation of the individual-based model illustrating disease transmission dynamics and the SEI model for each individual. Individuals transition among susceptible (S), exposed (E), and infectious (I) states following the indicated arrows. Transition rates are individual-dependent and may vary over time.}
  \label{fig:1}
\end{figure}

We investigated the dynamics of a one-day conference involving $N$ participants, denoted as $C_i$, where $i$ ranges from $1$ to $N$. Initially, only a handful of participants were assumed to be infected with the virus prior to the conference. Our analysis was confined to activities within the conference venue, as depicted in Figure \ref{fig:1}a. Transmission events may arise from interpersonal infections or environmental contamination in restroom facilities. To simulate disease spread, we employed an individual-based model that mirrors the behavior of each participant.

In our model, each of the $N$ individuals is characterized by their epidemic status, classified as susceptible (S),  exposed (E), or infectious (I). At the outset, all individuals are either susceptible or infectious, with susceptibility transitioning to exposure upon contact with the virus during the conference. Given the duration of a one-day conference, individuals typically do not progress from exposure to infectiousness within the same day. We adopted an SEI model to delineate the progression of epidemic statuses, whereby the infectious compartment remains stable throughout the conference duration (Figure \ref{fig:1}b).

Our model meticulously accounts for the activities of each participant as dictated by the conference itinerary, encompassing meetings, breaks, and meals. A representative conference schedule is outlined in Table \ref{tab:2}. During meetings, participants remain stationary, whereas, during breaks and meal times, they can move about and engage with others. Notably, restroom visits during these intervals pose potential opportunities for infection transmission. By simulating these individual activities, we estimated the potential number of individuals susceptible to infection following a one-day conference. 

\begin{table}[htbp]
\centering
\caption{The schedule of the conference}
\begin{tabular}{p{3.2cm}p{2cm}}
\hline
       Time     &  Schedule  \\
\hline
        $8:30-9:00$  & Rest time\\
        $9:00-10:00$   & Meeting  \\
        $10:00-10:30$    & Rest time \\
        $10:30-12:00$    & Meeting \\
        $12:00-14:00$    & Lunch break\\
        $14:00-16:00$   & Meeting   \\
\hline
\end{tabular}
\label{tab:2}
\end{table}

\subsection{Individual activities}

We analyzed each individual $C_i$, considering their spatial dynamics, represented by $\vec{r}_i$, and their epidemic status, denoted as $s_i$. The epidemic status is classified as susceptible (S), exposed (E), or infectious (I), with the transition from susceptible to exposed occurring subsequent to an infection event (to be detailed below).

To capture the movement of individuals across different locations $\vec{r}_i(t) = (x_i(t), y_i(t))$, we delineated two distinct modes: meeting and resting. During the meeting mode, participants are seated and stationary, adhering to pre-arranged seating arrangements. Typically, participants are assigned fixed seats for the duration of the conference. In cases where seat assignments vary, participants generally return to their original seats following breaks. Hence, we assumed each participant occupies a designated seat during the meeting sessions.

Conversely, during resting periods or the lunch break, participants can vacate their seats, traverse the venue, and engage in conversations. Movement patterns are simulated using a numerical two-dimensional random walk scheme, with parameters detailed in Table \ref{tab:1} in Appendix \ref{sec:A1}. Moreover, when individuals come into close proximity, they may momentarily pause their walk movement to engage in conversation before resuming their walk. Restroom visits were also accounted for during resting periods or the lunch break. Furthermore, participants may opt to remove their masks during the lunch break so that the mask-wearing rate is largely reduced. After the resting or lunch break period, most participants return to their original positions; however, a small fraction of participants, say 10\%, may change their seats. 

\subsection{Restroom timing}

The restroom serves as a focal point for potential infection transmission. To effectively model the infection events in this area, it is crucial to estimate the timing of participants' restroom visits. Typically, the waiting time for each participant to utilize the restroom follows an exponential distribution. However, since participants generally refrain from restroom visits during meetings, we must adjust the timing estimation to align with the conference schedule. 

In order to gauge the timing of restroom usage, we must ascertain, from the current time $t$, the probability of a participant using the restroom at time $t+a$. To this end, we define a probability function $p(a; t)$ such that
$p(a; t) \Delta a$ denotes the probability of a participant utilizing the restroom during the time interval $(t+a, t+a + \Delta a)$ from the first time, commencing from the current time $t$. 

Let $q(a; t)$ denote the likelihood of a participant refraining from using the restroom between the current time $t$ and $t + a$, while $f(a; t)$ represents the density function such that $f(a; t)\Delta a$ indicates the probability of a participant utilizing the restroom during the time interval $(t+a, t+a + \Delta a)$. Derived from $q(a; t)$ and $f(a; t)$, the probability density $p(a; t)$ is formulated as:
\begin{equation}
\label{eq:2.1}
p(a; t) = C(t)^{-1} f(a; t) q(a; t),\quad C(t) = \int_0^{+\infty} f(a; t) q(a; t)\mathrm{d} a.
\end{equation}

Detail derivations of the functions $q(a; t)$, $f(a; t)$, and $p(a; t)$ are given in Appendix \ref{sec:A1}.

To calculate the time interval that a participant uses the restroom, we utilize the distribution function $F(a; t)$, defined as
$$
F(a; t) = \int_0^a p(a'; t) \mathrm{d} a'.
$$
By employing $F(a; t)$ along with a uniform random number $s$ in the interval $[0, 1]$, we can solve the equation $s = F(a; t)$ to determine the time $a$ for a participant at the current moment $t$. The next visit to the restroom is scheduled at $t + a$. The numerical scheme to determine when a participant should next use the restroom, based on the current time $t$, is detailed in Appendix \ref{sec:A1}.

Utilizing the numerical scheme outlined in Appendix \ref{sec:A1}, we can approximate the timing of restroom usage for each participant based on their designated schedule $\{T_i\}$. Beginning from the current time, we compute the next restroom visit for each participant. The duration of restroom usage is determined by a random number specifically generated from the gamma distribution. Once an individual has finished using the restroom, we calculate their next scheduled visit and repeat the process accordingly.

\subsection{Virus transmission}

We investigated two scenarios for virus transmission: either an infectious individual transmits the virus to a susceptible individual or a susceptible individual contracts the virus from a contaminated restroom.

\subsubsection{Close contact infection}
Participants can move around during the resting or lunch breaks. If two individuals are within a certain distance of each other, known as the close contact distance ($\mathit{cdm}$), they are considered to be in close contact. Moreover, close contact infection may also occur during meetings if the distances between nearby seats are less than the close contact distance. If individuals $i$ and $j$ are in close contact, the duration of the contact is denoted as $t_{i,j}$. When a susceptible individual comes into close contact with an infectious individual, the likelihood of exposure depends on both the duration of contact and the measures taken to prevent transmission.

Let $\beta_{\max}$ denote the maximum probability of virus infection when a susceptible person comes into contact with an infected person for a sufficient amount of time. Infection of a susceptible individual $i$ by an infectious individual $j$ can be expressed as:
\begin{equation}
\beta_{i,j}(t_{i,j})=\beta_{\max}\times(1-e^{-t_{i,j}/\tau})\times M_{i,j}.
\end{equation}
Here, $M_{ij}$ signifies the protective effects of wearing masks; $t_{i,j}$ indicates the contact time between individual $i$ and individual $j$; $\tau$ represents a time constant that controls the increase of individual infection rate. The factor $M_{i,j}$ denotes a reduced factor for the infection rate when $i$ and/or $j$ wear a mask. We set the factor
\begin{equation}\label{eq:6}
M_{i,j}=\left\{
\begin{array}{ll}
     1.0, & i \ \mbox{and}\  j\, \mbox{are unmasked}\\
     0.33,      & i\ \mbox{is masked and}\ j\ \mbox{is unmasked}\\
     0.11,  & i\ \mbox{is unmasked and}\ j\ \mbox{is masked}\\
     0.017,   & i\ \mbox{and}\ j\ \mbox{are masked}
\end{array}\right.
\end{equation}
in our model simulation.

Now, consider the effect of vaccine protection, let $V_i$ represent the impact factor of vaccine (or recovery from infection) of an individual $i$. When a susceptible person $i$ contacts with multiple infectious individuals, the infection rate $\beta_{i}$ is given by
\begin{equation}
    \beta_{i}=\left(1-\prod_{i\ \mathrm{contacts\ with}\ j}(1-\beta_{i,j}(t_{i,j}))\right)\times(1-V_{i}).
\end{equation}
We set
\begin{equation}
V_{i}=\left\{
\begin{array}{lll}
     0, & i\ \mbox{has\, not\, been\, vaccinated}\\
     0.3,      & i\ \mbox{is vaccinated or recovered from infection}
\end{array}\right.
\end{equation}
in our model simulation.

\subsubsection{Infection from the contaminated restroom}

To formulate the infection rate of an individual from the contaminated restroom, we need to model the temporal changes of virus concentration and estimate the infection rate of infected objects to individuals in the restroom.

In the restroom, the virus concentration increases due to infectious individuals breathing and decreases through normal ventilation or regular disinfection. Thus, assuming each infectious individuals in the restroom releases the virus at a rate $\alpha$, and the virus is diluted in a time-dependent rate $b(t)$, changes in the virus concentration $X(t)$ can be expressed by the following differential equation:
\begin{equation}
   \frac{\mathrm{d} X}{\mathrm{d} t}=\frac{1}{V}\sum_{\mbox{infections}\ i}\alpha M_{i} \varepsilon_{i}(t) - b(t) X(t).
\end{equation}
Here $V$ represents the volume of the restroom, $\varepsilon_{i}(t)$ indicates whether the individual $i$ enters the restroom at time $t$, defined as
\begin{equation}\label{eq:10}
\varepsilon_{i}(t)=\left\{
\begin{array}{ll}
     1, & i\ \mbox{is at the restroom}\\
     0,      & \mbox{otherwise}.
\end{array}\right.
\end{equation}
The factor $M_{i}$ represents the protective factor of wearing a mask and is defined as:
\begin{equation}\label{eq:11}
  M_{i}=\left\{
\begin{array}{lll}
    1, & i\ \mbox{is unmasked}\\
     0.11,  & i\ \mbox{is infectious and masked}\\
     0.33, & i\ \mbox{is susceptible and masked.}
\end{array}\right.
\end{equation}
The dilution rate $b(t)$ represents the effects of normal ventilation and regular disinfection.

When a susceptible individual enters the restroom, the risk of getting infected depends on the amount of virus present. The infection rate increases as the concentration of the virus rises. Therefore, we can formulate the infection rate of a susceptible individual as:
\begin{equation}\label{eq:13}
  \beta_{i}(t) = \beta_{\max} \dfrac{(X(t)/X_0)^4}{1 + (X(t)/X_0)^4}M_{i}(1-V_{i}),
\end{equation}
where $M_i$ and $V_i$ are coefficients that represent the effects of wearing a mask and being vaccinated. In this context, a Hill function for $X(t)$ is presented to express how the concentration of the virus impacts the rate of infection, and $X_0$ denotes the virus concentration at the initial moment.

It is assumed that infection contaminates the environment through touch and droplets upon entering the restroom, potentially infecting susceptible individuals upon contact with contaminated objects such as door handles, toilet seats, and sinks. At this juncture, the infection status of susceptible is influenced by whether the infectious individuals are vaccinated and whether proper hand hygiene practices are followed. The probability of infection for a susceptible individual after using the restroom depends on the sequence of interactions with these objects. The ultimate comprehensive infection probability for susceptible individual $i$ is denoted as $\mu_i(t_0, s)$. For a detailed derivation, please refer to Appendix \ref{sec:A3}.

Utilizing the provided infection rate, when an individual uses the restroom between the time period of $t = t_0$ and $t = t_0 + s$, the likelihood of being infected by the virus and contaminated objects within the restroom is represented by:
\begin{equation}
\label{I_solution7}
 p_{i}(t_0, s) = 1-\exp\left[{-\int_{t_{0}}^{t_{0}+s}\beta_{i}(s)ds}\right] \times (1 - \mu_i(t_0, s))
\end{equation}
Please see Appendix \ref{sec:A4} for a detailed derivation. This allows us to determine whether an individual becomes infected after finishing restroom use.

In the simulations, our primary focus is on highly infectious viruses, such as the Omicron strain of SARS-CoV-2. Therefore, we assumed the maximum infected probability $\beta_{\max} = 0.9$.

\section{Results}

We employed the individual-based model to analyze disease transmission dynamics during a one-day conference. Our objective is to assess the extent of infections among participants following the conference, given a few initially infected individuals. For our simulations, we utilized the schedule outlined in Table \ref{tab:2}, with the conference commencing at 9:00 am and all participants arriving 30 minutes prior. Throughout the simulations described below, we considered a scenario with $N=250$ participants and initially $3$ infectious individuals.

\subsection{Infections without control measures}

\begin{figure}[htbp]
\centering
\begin{center}
\includegraphics[width=12cm]{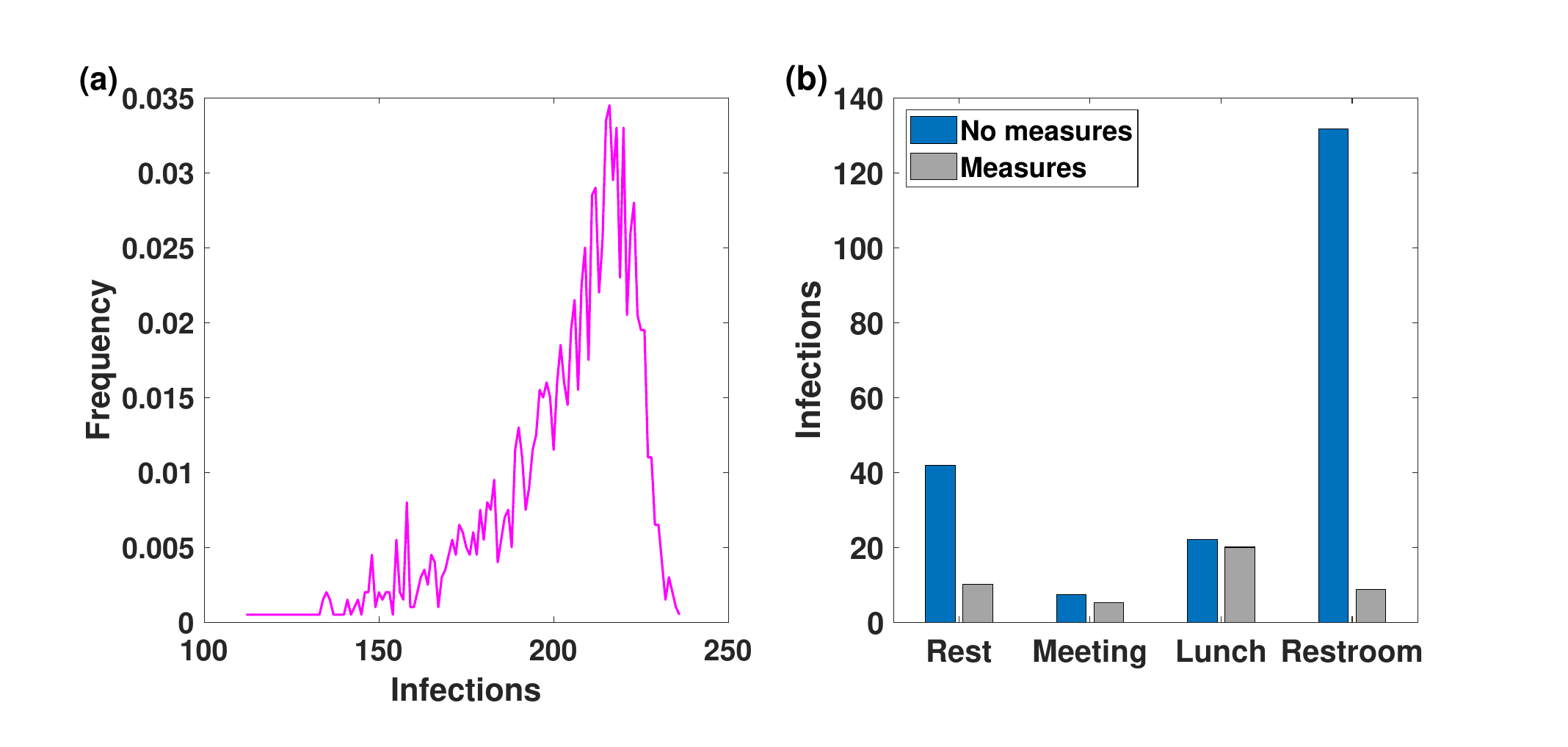}
\end{center}
\caption{Comparison of infections number with and without prevention and control measures. (a) Cumulative average infections without prevention and control measures. The horizontal axis represents the number of infections, while the vertical axis denotes the frequency of infection occurrences. (b) Distribution of Infections across various time periods. Color bars indicate the number of infections during resting, meetings, lunch breaks, and restroom usage, with or without prevention and control measures, respectively. All results were derived from $2000$ independent runs.}
\label{fig:3}
\end{figure}

Initially, we explored scenarios where no control measures, such as mask-wearing, vaccination, or restroom disinfection (ventilation), were implemented. Our model simulations aimed to estimate the final tally of infections by the end of the one-day conference. In the absence of control measures, simulations revealed that the number of infections ranged from $100$ to $240$, with an average of $210$ ($84\%$). Figure \ref{fig:3}a illustrates the frequency distribution of infection numbers. 

To delver deeper into the infection dynamics, we scrutinized the average infection counts during distinct time frames, including resting periods, meetings, lunch breaks, and restroom use (Figure \ref{fig:3}b). Our analyses unveiled that few infections occurred during meetings, whereas most infection events transpired during lunch breaks and rest periods. Notably, a significant number of infections were also linked to restroom usage. These findings underscore the necessity of implementing control measures during breaks and resting periods, alongside adopting measures such as restroom disinfection (ventilation) to mitigate the spread of infections.

\subsection{Effects of different control measures}

\begin{figure}[htbp]
\centering
\includegraphics[width=14cm]{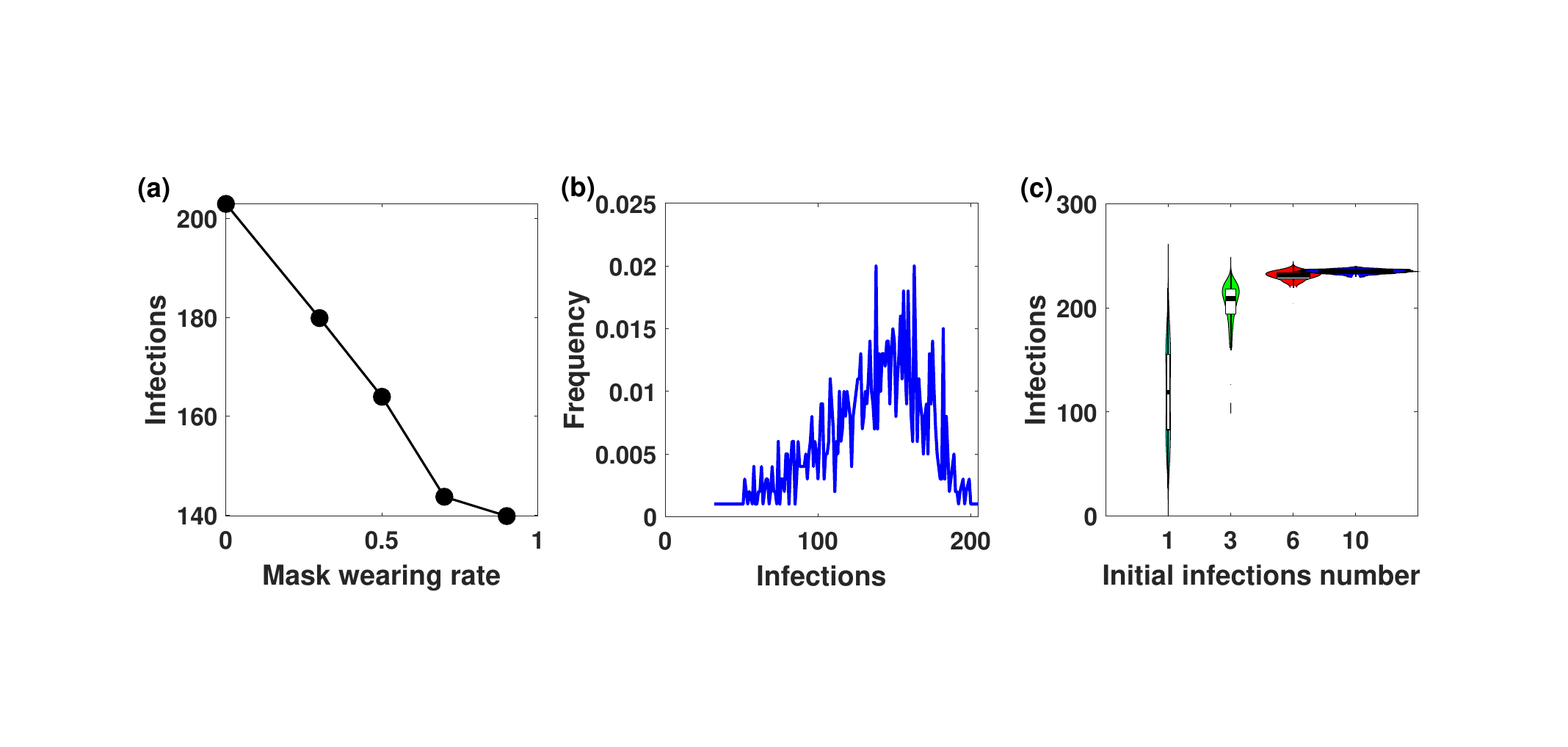}
\caption{Impact of {\textbf{(a)}} mask wearing rate, {\textbf{(b)}} wearing masks of all initial infections, and {\textbf{(c)}} initial infections number on the epidemic dynamics. All results were derived from $1000$ independent runs.}
\label{fig:4}
\end{figure}

We proceed to evaluate the effectiveness of various control measures. Our primary recommendation involves enforcing mask-wearing among participants, except during lunch breaks. To gauge the efficacy of this measure, we conducted simulations by varying the mask-wearing rate among participants while holding other parameters unchanged. Figure \ref{fig:4}a illustrates the relationship between the average number of infections and the percentage of participants wearing masks. Our findings indicate a significant reduction in the number of infections with an increase in the mask-wearing rate. Notably, when 90\% of participants wear masks, the average number of infections decreases to 105 - a stark contrast to the 200 infections recorded when no one wears a mask. This underscores the substantial protective effect of increasing mask-wearing rates, effectively minimizing the risk of infection among susceptible individuals and playing a pivotal role in epidemic control.

One strategy to curb the spread of infectious diseases at conferences involves mandating masks for all initially infected individuals. Through nucleic acid testing (NAT), participants testing positive can be identified, albeit they may need to attend the conference for various reasons. Our approach entails requiring these infections to wear masks to mitigate the final number of infections. Figure \ref{fig:4}b illustrates that infection numbers mostly range between 60 and 150, with the highest concentration falling between 100 and 120. Notably, the infection count is lower compared to scenarios without any prevention or control measures, indicating the efficacy of this approach in curtailing the final epidemic size. 

Another approach to prevent the spread of infectious disease at conferences involves preventing infected individuals from attending altogether. In the case of COVID-19, this can be achieved through NAT to reduce the number of initial infections. To assess the effectiveness of NAT, we varied the initial infection numbers $(1, 3, 6, 10)$ and assumed a 60\% mask-wearing rate among participants. Our results reveal a significant reduction in the final size following the conference with a decrease in the number of initial infections (Figure \ref{fig:4}c). Specifically, there exists a strong correlation between the initial infection number and the final epidemic size.

\subsection{Effects of adjusting the meeting schedule}

\begin{figure}[htbp]
\centering
\begin{center}
\includegraphics[width=12cm]{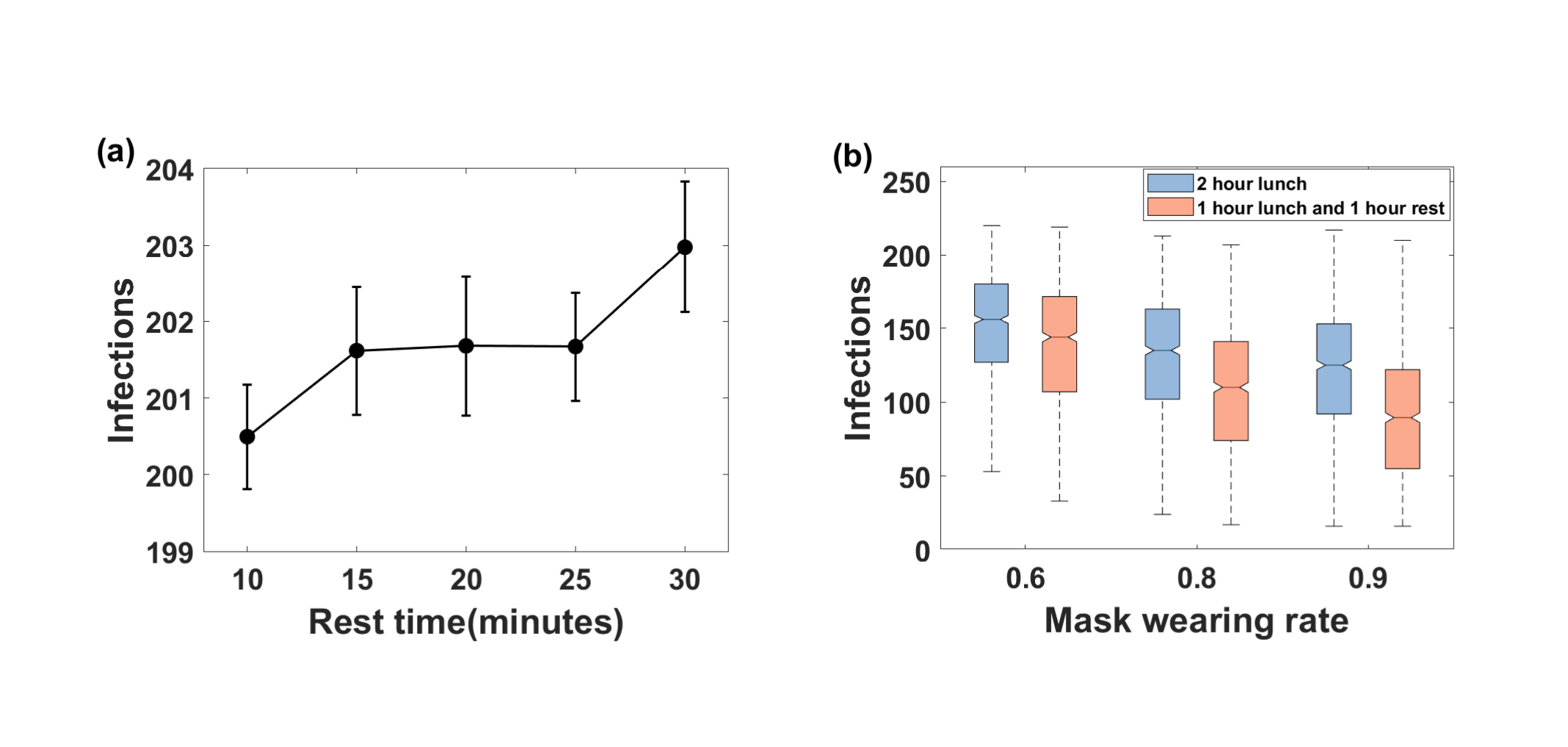}
\end{center}
\caption{Impact of {\textbf{(a)}} shortening rest periods in the meeting schedule and {\textbf{(b)}} adjusting lunch time on infection numbers. The rest durations explored were $10, 15, 20, 25$, and $30$ minutes, respectively. Adjustments included splitting a 2-hour lunch into a 1-hour rest and a 1-hour lunch, with varying mask-wearing rates among individuals during the resting period. Results were based on $1000$ independent runs.
}
\label{fig:5}
\end{figure}

Our findings indicate that most infections occur during rest periods or lunch breaks. To minimize the risk of infections, adjustments to the meeting schedule, such as reducing the duration of these breaks, were considered. We began by analyzing the impact of shortening the rest periods, examining resting intervals of 10, 15, 20, 25, or 30 minutes. Figure \ref{fig:5}a shows that reducing the rest time from 30 to 15 minutes does not significantly impact the number of infections. Therefore, merely shortening rest periods may not effectively reduce the overall number of infections.

Further exploration involved adjusting the structure of lunch breaks, such as dividing a 2-hour lunch into a 1-hour lunch and a 1-hour rest period. The adjustment replaced a 1-hour lunch break with a resting period so that participants would wear the mask during the resting period. According to Figure \ref{fig:5}b, the number of infections decreases as the mask-wearing rate increases because of the shortening of the lunch break. The configuration of a 1-hour lunch followed by a 1-hour rest period results in fewer infections compared to a continuous 2-hour lunch period. These outcomes suggest that restructuring lunch and rest times can effectively reduce infection numbers, contributing positively to epidemic control measures.

\subsection{Impact of multiple control measures on epidemic evolution}

In our analysis, we considered various prevention and control measures individually to understand their impact on the epidemic. However, to comprehensively assess the effectiveness of epidemic prevention and control, we conducted a comprehensive analysis of multiple control measures. These measures, detailed in Table \ref{tab:3}, include participant mask-wearing, schedule adjustments, restroom management, and reducing initial infections through NAT.

\begin{figure}[htbp]
\centering
\begin{center}
\includegraphics[width=12cm]{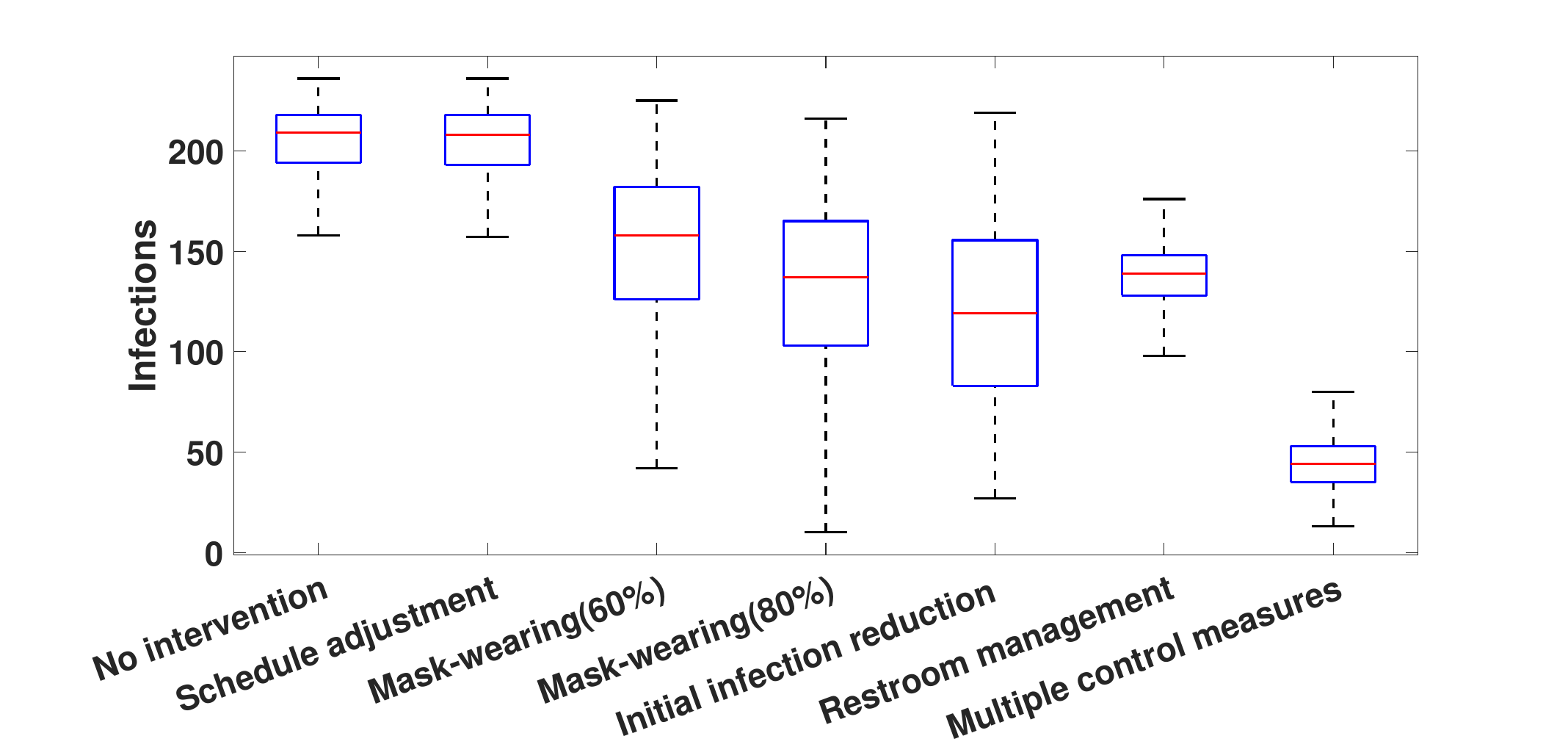}
\end{center}
\caption{Effect of multiple control measures on infections. These measures include no intervention, participant mask-wearing (60\% or 80\%), schedule adjustments, initial infection reduction, restroom management, and a combination of multiple measures. Details of each measure are provided in Table \ref{tab:3}. All results were based on $1000$ independent simulation runs. 
}
\label{fig:6}
\end{figure}

To compare the efficacy of these measures, Figure \ref{fig:6} represents the final infection numbers under different scenarios: no control measures, single control measures, and a combination of multiple measures. The results demonstrate that implementing any single control measure, such as wearing masks or schedule adjustments, leads to a reduction in infection compared to no intervention. However, the most significant infection reduction occurs when multiple control measures are combined. 

From the simulation results in Figure \ref{fig:6}, we observe that reducing the initial infection number to 1, maintaining a 60\% mask-wearing rate among participants, adjusting the lunch schedule to a 1-hour lunch and a 1-hour rest, and implementing restroom disinfection and ventilation can reduce the final infection count from 200 to 100, representing a 50\% reduction.

\begin{table}[htbp]
\centering
\caption{Prevention and control measures}
\begin{tabular}{p{5.2cm}p{9cm}}
\hline
       Types of control measures     &  Specific control measures  \\
\hline
        Participant mask-wearing   & Mask-wearing rate of 60\% or 80\% among participant  \\
        Schedule adjustment    &  Adjustment of the 2-hour lunch break to 1-hour lunch and a 1-hour rest \\
        Restroom management   & Regular disinfection and ventilation of restrooms \\
        Initial infection reduction &  Reduction of initial infectious to 1 \\
        Multiple control measures    & Combination of participant mask-wearing (at 60\% rate), schedule adjustments, restroom management, and initial infection reduction\\
\hline
\end{tabular}
\label{tab:3}
\end{table}

\section{Conclusion}

Our study presents a comprehensive computational model designed to simulate the spread of infectious disease during one-day conferences at an individual level. By categorizing conference schedules into various activities such as meetings, resting periods, and lunch breaks, we accurately capture participant movements and the potential for infection within a short timeframe. Notably, our model considers the transmission dynamics of highly infectious viruses like the Omicron strain of SARS-CoV-2.

Our results underscore the importance of implementing effective prevention and control measures to mitigate the spread of infectious diseases during conferences. Without any interventions, our simulations show a high rate of infection, with nearly 84\% of participants becoming infected by the end of the conference day. However, implementing single control measures, such as mask-wearing, reducing initial infections, and adjusting meeting schedules, can effectively reduce the final infection count.

Of particular significance is the finding that the simultaneous implementation of multiple control measures yields the most substantial reduction in infections. By combining measures such as participant mask-wearing, schedule adjustments,  restroom management, and initial infection reduction, the number of infections can be significantly lowered, potentially to a very low level.

The computational model proposed in this study serves as a valuable quantitative assessment and prediction tool for studying the dynamics of epidemic evolution under various prevention and control conditions. It enables the prediction of the potential impact of different prevention and control measures under different scenarios, providing insights for policymakers and conference organizers alike.

 Our findings offer practical guidance for conference organizers in minimizing the risk of disease transmission at their events. Furthermore, our research contributes to the quantitative assessment of prevention and control measures against COVD-19 and similar infectious diseases, providing a foundation for future studies in this area.

\section*{Appendix}

\setcounter{section}{0}
\renewcommand{\thesection}{A.\arabic{section}}

\setcounter{figure}{0}
\renewcommand{\thefigure}{A.\arabic{figure}}

\setcounter{table}{0}
\renewcommand{\thetable}{A.\arabic{table}}

\setcounter{equation}{0}
\renewcommand{\theequation}{A.\arabic{equation}}

\section{Parameters of the random walk}
\label{sec:A1}

Table \ref{tab:1} listed the parameters used to simulate the random walk of individuals.

\begin{table}[htbp]
\centering
\label{tab:1}
\caption{Parameters of the random walk}
\begin{tabular}{p{2.2cm}p{5.2cm}p{3.2cm}p{2.4cm}}
\hline
       Parameter  &  Description    &  Value$^{(a)}$ & Unit \\
\hline
        $\mathit{CR}$  & the conversation rate of two individuals  &  $\mathrm{uniform}(0, 1)$ &   $\mathrm{sec}^{-1}$ \\
        $\mathit{TR}$   & upper limit of the conversation rate of two individuals  &  0.075 &   $\mathrm{sec}^{-1}$ \\
        $\mathit{NR}$   &  movement probability of individual  & $\mathrm{uniform}(0, 1)$  &   $\mathrm{sec}^{-1}$ \\
        $\mathit{MR}$   & upper limit of the movement probability &  0.175 &  $\mathrm{sec}^{-1}$\\
        $\mathit{talklimit}$   & upper limit of individual conversation time   &  $\mathrm{Gamma}(\mathit{ta}, \mathit{tb})$ &   $\mathrm{sec}$ \\
        $\mathit{ta}$   & a parameter of limited conversation time  &  80 &  $\mathrm{sec}$\\
        $\mathit{tb}$   & another parameter of limited conversation time  &  6 &  $\mathrm{sec}$\\
        $\mathit{cd}$     &  close conversation distance & 0.5 &  $\mathrm{m}$ \\
        $\mathit{cdm}$  & close contact distance  & 2 &  $\mathrm{m}$ \\
        $\mathit{sa}$  & the abscissa of the initial coordinate position  & 2 &  $\mathrm{m}$ \\
        $\mathit{sb}$  & the ordinate of the initial coordinate position  & 2 &  $\mathrm{m}$ \\
\hline
\end{tabular}
\begin{minipage}{14cm}
\begin{enumerate}
\item[$^{(a)}$] $\mathrm{uniform}(0, 1)$ means a random number with uniform distribution in $[0, 1]$, $\mathrm{Gamma}(\mathit{ta}, \mathit{tb})$ means a gamma distribution random number with parameter $\mathit{ta}$ and $\mathit{tb}$.   
\end{enumerate}
\end{minipage}
\end{table}

\section{Timing of the next visit to the restroom}
\label{sec:A2}

The function $p(a; t)$ represents the probability density such that
$p(a; t) \Delta a$ denotes the probability of a participant utilizing the restroom during the time interval $(t+a, t+a + \Delta a)$ from the first time, commencing from the current time $t$. The function $q(a; t)$ denotes the likelihood of a participant refraining from using the restroom between the current time $t$ and $t + a$, and $f(a; t)$ represents the density function such that $f(a; t)\Delta a$ indicates the probability of a participant utilizing the restroom during the time interval $(t+a, t+a + \Delta a)$. We have
\begin{equation}
p(a; t) = C(t)^{-1} f(a; t) q(a; t), \quad C(t) = \int_0^{+\infty} f(a; t) q(a; t) \mathrm{d} a.
\end{equation}

\subsection{The function $q(a; t)$ and $f(a; t)$}

\begin{figure}[htbp]
\centering
\includegraphics[width=8cm]{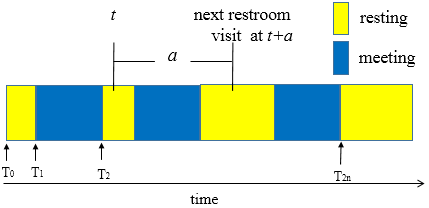}
\caption{Estimating the timing of restroom usage.}
\label{fig:rest}
\end{figure}

To estimate the functions $q(a; t)$ and $f(a; t)$, we introduce the following assumptions:
\begin{enumerate}
\item[(1)] Each participant typically visits the restroom with a uniform frequency $f_0$, implying that the probability of a restroom visit within a short time interval $\Delta t$ is $f_0 \Delta t$;
\item[(2)] The conference schedule time points (depicted in Figure \ref{fig:rest}) are denoted as:
$$
T = (T_0, T_1, T_2, \cdots, T_{2n}),
$$
where $T_0$ signifies the commencement of the conference, the intervals $[T_{2k}, T_{2k+1}]$ represent rest or lunch breaks, and $[T_{2k+1}, T_{2(k+1)}]$ signify meeting periods. Participants do not visit the restroom during meetings;
\item[(3)] All participants depart from the conference venue at $T_{2n}$;
\item[(4)] Restroom visits by participants are independent events.
\end{enumerate}

Firstly, we assumed that participants are presently in a break period $t \in [T_{2k}, T_{2k+1}]$, disregarding scheduled meetings, allowing participants to utilize the restroom at their convenience freely. Consequently, we have:
\begin{equation}
\label{eq:1}
q(a; t) = e^{-f_0 a}.
\end{equation}

Now, we account for the impact of the meeting schedule:
\begin{enumerate}
\item[(1)] It the time point $t+a$ falls within the resting interval $[T_{2k}, T_{2k+1}]$, \textit{i.e.}, $0 < a < T_{2k+1} - t$, we maintain:
\begin{equation}
\label{eq:2}
q(a; t) = e^{-f_0 a}.
\end{equation}
\item[(2)] When the time point $t + a$ aligns with the meeting period, \textit{i.e.}, $t + a \in [T_{2j+1}, T_{2(j+1)}]$,  or $a \in [T_{2j +1} - t, T_{2(j+1)} - t]$ ($j\geq k$), participants abstain from restroom visits during meetings. Therefore:
\begin{equation}
\label{eq:3}
q(a; t) = q(T_{2j+1} -t; t).
\end{equation}
\item[(3)] If the time point $t + a$ coincides with the resting period, \textit{i.e.}, $t + a\in [T_{2j}, T_{2j+1}]\ (j\geq k+1)$, participants may exhibit a heightened need for restroom visit due to lack of access during meetings. Let $\hat{f}_j$ denote the adjusted restroom visit frequency during $t + a\in [T_{2j}, T_{2j+1}]$. Noting that $q(a; t)|_{a = T_{2j}- t} = q(T_{2j-1} - t; t)$, we express:
\begin{equation}
\label{eq:4}
q(a; t) = q(T_{2j-1}-t; t) e^{-\hat{f}_j(a + t - T_{2j})},\quad a\in [T_{2j}-t, T_{2j+1}-t].
\end{equation}
The adjusted frequency $\hat{f}_j$ can be determined by assuming participants have an equivalent probability of not using the restroom during $[t, t+ T_{2j+1}]$ as they would without meetings. Hence, comparing \eqref{eq:1} with \eqref{eq:4}, we derive:
$$
e^{-f_0(T_{2j+1} - t)} = q(T_{2j-1}; t) e^{-\hat{f}_j (T_{2j+1} - T_{2j})},
$$
yielding:
$$
\hat{f}_j = \dfrac{f_0 (T_{2j+1} - t) + \ln q(T_{2j-1} - t; t)}{T_{2j+1} -T_{2j}}.
$$
Moreover, following the frequency adjustment, the probability $q(T_{2j-1}-t; t)$ equates to that without meetings, $i.e.$,
$$
q(T_{2j-1}-t; t) = e^{-f_0 (T_{2j-1} -t)}.
$$
Thus, the adjusted frequency becomes:
\begin{equation}
\label{eq:5}
\hat{f}_j = \dfrac{T_{2j+1} - T_{2j-1}}{T_{2j+1} - T_{2j}} f_0,
\end{equation}
resulting in:
\begin{equation}
q(a; t) = e^{-f_0(T_{2j-1} - t) + \hat{f}_j(T_{2j} - t)} e^{-\hat{f}_j a}.
\end{equation}
\end{enumerate}

In summary, we obtain (where $t\in (T_{2k}, T_{2k+1})$):
\begin{equation}
\label{eq:7}
q(a; t) = \left\{
\begin{array}{ll}
e^{-f_0 a},\quad & t < a + t \leq T_{2k+1}\\
e^{-f_0(T_{2j+1} - t)}, \quad &  T_{2j+1} \leq a + t \leq T_{2(j+1)},\ j\geq k\\
e^{-f_0(T_{2j-1} - t) + \hat{f}_j(T_{2j} - t)} e^{-\hat{f}_j a},\quad & t_{2j} \leq a + t \leq T_{2j+1},\ j\geq k+1.
\end{array}
\right.
\end{equation}
Here, $T_{2n+1} = +\infty$, thus $\hat{f}_n = f_0$.

Let $f(a; t)\Delta a$ represent the probability of a participant using the restroom during the time interval $(t+a, t+a+\Delta a)$, then:
\begin{equation}
\label{eq:8}
f(a; t) = \left\{
\begin{array}{ll}
f_0,\quad & t < a + t \leq T_{2k+1}\\
0, \quad &  T_{2j+1} \leq a + t \leq T_{2(j+1)},\ j\geq k\\
\hat{f}_j,\quad & T_{2j} \leq a+t \leq T_{2j+1},\ j\geq k+1.
\end{array}
\right.
\end{equation}

\subsection{The fuction $p(a; t)$}

Now, let's evaluate $C(t)$, which is given by
\begin{eqnarray*}
C(t) &=& \int_0^{+\infty} f(a; t) q(a; t) \mathrm{d} a\\
&=&\int_0^{T_{2k+1} - t} f_0 e^{-f_0 a} \mathrm{d} a\\
&&{} + \sum_{j=k+1}^{n-1} \int_{T_{2j} - t}^{T_{2j+1} - t} \hat{f}_j e^{-f_0 (T_{2j-1} - t) + \hat{f}_j(T_{2j} - t)} e^{-\hat{f}_j a} \mathrm{d} a\\
&&{} + \int_{T_{2n} - t}^{+\infty} \hat{f}_n e^{-f_0 (T_{2n-1} - t) + \hat{f}_n(T_{2n} - t)} e^{-\hat{f}_n a} \mathrm{d} a\\
&=& 1 - e^{-f_0(T_{2k+1} - t)} + \sum_{j = k+1}^{n-1} e^{-f_0 (T_{2j - 1} - t)} (1 - e^{-\hat{f}_j (T_{2j+1} - T_{2j})}) + e^{-f_0(T_{2n-1} - t)}.
\end{eqnarray*}
The probability $p(a, t)$ is then given by
\begin{equation}
\label{eq:9}
\begin{aligned}
p(a; t) &= C(t)^{-1} f(a; t) q(a; t)\\
&=C(t)^{-1}\times \left\{
\begin{array}{ll}
f_0 e^{-f_0 a},\quad & t < a + t \leq T_{2k+1}\\
0, \quad &  T_{2j+1} \leq a + t \leq T_{2(j+1)},\ j\geq k\\
\hat{f}_j e^{-f_0(T_{2j-1} - t) + \hat{f}_j(T_{2j} - t)} e^{-\hat{f}_j a},\quad & T_{2j} \leq a+t \leq T_{2j+1},\ j\geq k+1.
\end{array}
\right.
\end{aligned}
\end{equation}

\subsection{The function $F(a; t)$ and the numerical scheme to determine the next time to visit the restroom}

\begin{figure}[htbp]
\centering
\includegraphics[width=8cm]{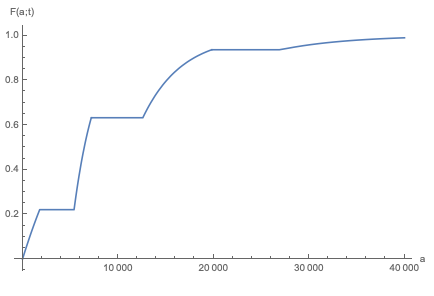}
\caption{Plot of the function $F(a; t)$}
\label{fig:A.2}
\end{figure}

To calculate the time interval that a participant uses the restroom, we utilize the distribution function $F(a; t)$, defined as
$$
F(a; t) = \int_0^a p(a'; t) \mathrm{d} a'.
$$
The graph of the function $F(a; t)$ is shown in Figure \ref{fig:A.2}. By employing $F(a; t)$ along with a uniform random number $s$ in the interval $[0, 1]$, we solve the equation $s = F(a; t)$ to determine the time $a$ for a participant at the current moment $t$. The next visit to the restroom is scheduled at $t + a$.

Let's define
$$
\varphi_i(a, t) = \left\{
\begin{array}{ll}
1 - e^{-f_0 a},\quad & i = k\\
e^{-f_0(T_{2i - 1} - t)} (1 - e^{\hat{f}_i(a + t - T_{2i})}),\quad & (k < i \leq n)
\end{array}
\right.
$$
Then $C(t)$ can be expressed as (we note $t_{2n+1} = \infty$)
$$
C(t) = \sum_{i=k}^n \varphi_i(T_{2i + 1} - t; t).
$$
A detailed calculation leads to:
$$
F(a, t) = \left\{
\begin{array}{ll}
\displaystyle C(t)^{-1} \varphi_k(a; t),\quad & 0 \leq a < T_{2k+1} -t,\\
\displaystyle C(t)^{-1} \sum_{i=k}^j \varphi_i(T_{2i+1} - t; t), & T_{2k+1} - t \leq a < T_{2(k+1)} - t,\\
\displaystyle C(t)^{-1} \left(\sum_{i=k}^{j-1} \varphi_i(T_{2i + 1} - t; t) + \varphi_j(a; t)\right),\quad & T_{2j} - t \leq a < T_{2j+1} - t\ (k+1 \leq j \leq n).
\end{array}
\right.
$$

Here is a numerical scheme to determine when a participant should next use the restroom, based on the current time $t$:\\

\noindent\textbf{Input:} The current time $t$, the schedule $\{T_i\}$\\
\textbf{Output:} The time of the next restroom visit $t + a$. 
\begin{enumerate}[1.]
\item Find the value of $k$ such that $T_{2k} < t <T_{2k+1}$.
\item Calculate $A_{i}=\varphi_{i}(T_{2i+1}-t;t), (i=k,k+1,\ldots,n)$.
\item Calculate
$$C=\sum_{i=k}^{n}A_{i}.$$
\item Generate a uniform random number $s\in [0, 1]$.
\item Find the value of $j$ such that $k\leq j\leq n$ and
$$\frac{1}{C}\sum_{i=k}^{j-1}A_{i}<s<\frac{1}{C(t)}\sum_{i=k}^{j}A_{i},$$
then calculate
$$s_{0}=s-\frac{1}{C}\sum_{i=k}^{j-1}A_{i}$$
\item Solve the equation
$$\varphi_{j}(a;t)=s_{0}$$
for $a$, and the time of the next restroom visit is given by $t+a$. Explicitly,
\begin{equation}
     t+a = \left\{
\begin{array}{ll}
     t -{f_0}^{-1}\ln(1-s_{0}),\quad  & j=k,\\
     T_{2j} - {\hat{f}_j}^{-1}\ln(1-s_{0}e^{f_0(T_{2j-1}-t)}),\quad  & j>k.
\end{array}\right.
\end{equation}
\end{enumerate}

Utilizing the above numerical scheme, we can approximate the timing of restroom usage for each participant based on their designated schedule $\{T_i\}$. 

\section{Environmental infection rate of restroom}
\label{sec:A3}

Here, we calculate the dynamics of environmental viral load in the restroom during the conference and subsequently estimate the environmental infection rate. Initially, our focus lies on assessing virus transmission within the restroom environment. Consequently, we primarily consider the dynamics of viral load, accounting for both its increase and decrease due to regular ventilation and disinfection procedures. Moreover, individuals using the restroom must be categorized as either infectious or susceptible. If deemed infectious, the total viral load in the environment at that time requires calculation. Conversely, for susceptible individuals, the environmental infection rate needs to be determined.

Subsequently, we delve into the transmission of the virus from contaminated objects to individuals. We presume that infectious individuals contaminate the environment through touch and droplets upon entering the restroom, posing a risk of infection to susceptible individuals upon contact with contaminated surfaces. The infection risk for susceptibles is influenced by factors such as vaccination status and hand hygiene practices. Notably, we assumed that transmission via touch is primarily through direct contact rather than airborne transmission, thus rendering mask-wearing rates irrelevant in this context.

First, we introduce several key notations used in our derivation:
\begin{itemize}
\item $V$: The volume of the restroom.
\item $X(t)$: The virus load of the restroom at the current moment. Initially, $X=0$. The viral load increases as infectious individuals enter the restroom, and decreases due to regular ventilation and disinfection.
\item $\alpha$: The rate infectious individuals release the virus.
\item $b(t)$: The reduction rate of virus at time $t$.
\item $\beta_{i}(t)$: The environmental infection rate of susceptible $i$ in the restroom from time $t$, so that $\beta_i(t) \Delta t$ indicates the probability that susceptible $i$ was not infected before time $t$, but was infected within $[t, t+ \Delta t]$.
\item $\beta_{\max}$: The maximum infection rate of the virus to individuals.
\item $\beta_{\max(i)}$: The maximum infection rate of the virus to individual $i$.
\item $S_0$: The basic infection rate of infected objects to individuals in the restroom (including door handles, stalls, and sinks).
\item $d(t)$: Indicates whether the restroom door handle of the restroom is infected.
\item $h_{j}(t)$: Indicates whether the $j$th stall in the restroom is infected. 
\item $\mu_i$: The probability that susceptible individual $i$ in the restroom is infected by infected objects.
\item $PW$: The probability that individuals go to the restroom and wash their hands.
\item $M_{i}$:  The protective effect of wearing a mask for individual $i$:
\begin{equation}
  M_{i}=\left\{
\begin{array}{lll}
     1, & i\, \mbox{is unmasked}\\
     0.11, & i\, \mbox{is infectious and masked}\\
     0.33,  & i\, \mbox{is susceptible and masked}\\
\end{array}\right.
\end{equation}
\item $V_{i}$: The impact factor of vaccine for susceptible individuals:
\begin{equation}
\label{tranf20}
V_{i}=\left\{
\begin{array}{lll}
     0, & i,\,  \mbox{do not has vaccine}\\
     0.3,      & i\, \mbox{is vaccinated, or recovered from infection} \\
\end{array}\right.
\end{equation}
\item $\varepsilon_{i}$: Indicates whether individual $i$ enters the restroom at time $t$:
\begin{equation}\label{tranf20}
\varepsilon_{i}(t)=\left\{
\begin{array}{lll}
     1, & i\, \mbox{is infectious and enters the environment}\\
     0,      & \mbox{otherwise}\\
\end{array}\right.
\end{equation}
\end{itemize}

For each object in the restroom, a status variable indicates its infection status. When the infectious enters the restroom and comes into contact with the object, if the infection status of the object becomes $1$, it means the object is infected. The infection status of all objects changes to $0$ after the next disinfection cycle. To implement this, we assign a corresponding infection state variable for each object (including the door, each stall in the restroom, and each sink). Then, based on the contact between the object and the infectious individual, we set the infection state variable accordingly. 

Using this process, we can determine the infection situation $d(t)$ of the door handle and the infection situation $h_j(t)$ for the $j$th stall in the restroom. The corresponding expressions for these are as follows: the infection situation of the door handle $d(t)$ is defined as:
$$d(t) = \left\{
\begin{array}{ll}
1,  & \mbox{if the handle is infected}, T_0 < t < T_1,  \mbox{where}\ T_0\ \mbox{is the time of the first infectious} \\
& \mbox{indivituals entering the restroom (assuming the individual will definitely touch the door)},\\
& T_1\, \mbox{is the time of the first disinfection from}\ T_0\\
0,& \mbox{if the door is not infected, at other times}
\end{array}
\right.
$$

For restrooms without a restroom door, we $d(t) = 0$ for simplicity. The infection status of the $j$th stall $h_{j}(t)$ is expressed as:
$$
h_{j}(t) = \left\{
\begin{array}{ll}
1, & \mbox{the}\, j\mathrm{th}\, \mbox{stall is infected}, T_{0,j} < t < T_1, \mbox{where}\ T_0\ \mbox{is the time when the infectious}\\
& \mbox{individuals first enter the}\  j\mathrm{th}\ \mbox{stall},\ T_1\  \mbox{is the first disinfection time from}\ T_0\\
0,& \mbox{if the}\, j\mathrm{th}\ \mbox{stall is not infected, at other times}
\end{array}
\right.
$$
We did not consider the sink's infection here, assuming hand washing can effectively eliminate virus transmission. However, in practice, individuals may become infected through sinks, albeit with low probability.

It's important to note that the function of the infection status is not easily described by a unified mathematical formula. In practical calculations, the infection status of each object should be determined based on the specific calculation process rather than relying solely on mathematical formulas.

Next, let's examine the probability $\mu_i$ that a susceptible individual is infected by contaminated objects in the restroom. Suppose individuals enter the restroom at time $t_0$ and stay there for a duration of $s$. We assume the order of object contact is: restroom door $\rightarrow$ stall $j$ $\rightarrow$ restroom door $\rightarrow$ sink. Then, individuals come into contact with the restroom door and the stall at $t = t_0$, and with the restroom door and the sink at $t = t_0 + s$.

Let $w_i$ indicate whether individual $i$ washes hands. Individual $i$ washes hands with probability $PW$ and does not wash hands with probability $1-PW$. Based on our assumptions, if the basic infection rate of object contact is $S_0$, the probability of individual $i$ becoming infected after contact with the door at time $t$ is:
$$S_0 d(t) (1 -  V_i).$$
Therefore, the probability of individual $i$ remaining uninfected after contact with the door is:
$$1 - S_0 d(t) (1 -  V_i).$$
Similarly, the probability of individual $i$ remaining uninfected after contacting the stall $j$ at time $t$ is:
$$1 - S_0 h_j(t) (1-V_i).$$
Considering these situations, the probability that individual $i$ remains uninfected after contacting the outside door and the stall $j$ at time $t_0$, and remains uninfected after contacting the restroom door at time $t_0 + s$ is:
$$(1 - S_0 d(t_0) (1 -  V_i))(1 - S_0 h_j(t_0) (1-V_i))(1 - S_0 d(t_0 + s) (1 -  V_i)).$$
The probability of individual $i$ becoming infected is then:
$$1 - (1 - S_0 d(t_0) (1 -  V_i))(1 - S_0 h_j(t_0) (1-V_i))(1 - S_0 d(t_0 + s) (1 -  V_i)).$$
Considering whether individuals wash their hands, the final comprehensive infection probability is:
\begin{equation}
\label{eq:28}
\mu_i(t_0, s) = (1- w_i)(1 - (1 - S_0 d(t_0) (1 -  V_i))(1 - S_0 h_j(t_0) (1-V_i))(1 - S_0 d(t_0 + s) (1 -  V_i))).
\end{equation}
Equation \eqref{eq:28} provides the infection of individual $i$. However, it's important to emphasize that the infection rate should not be solely calculated using this formula in practical calculations. Instead, the infection situation should be determined by simulating the process of an individual using the restroom.

\section{The infection rate of individuals after using the restroom}
\label{sec:A4}

The following method outlines the calculation of the probability of each individual being infected after using the restroom during a rest period. The primary approach is to compute the probability of individuals remaining uninfected after using the restroom and then deduce the probability of them becoming infected. In our model, we neglect scenarios involving individuals conversing in the restroom. Consequently, there are two primary pathways through which susceptible individuals may become infected in the restroom: direct exposure to viruses present in the environment and contact with contaminated objects. Therefore, the probability of individuals contracting an infection in the restroom depends on factors such as the viral concentration in the environment, the duration of the individual's restroom visit, and the likelihood of contact with contaminated surfaces.

Assuming an individual $i$ enters the restroom at time $t_0$, we computed the probability of that individual becoming infected during their stay in the restroom for a duration of $s$. 

\begin{figure}[htbp]
\centering
\includegraphics[width=10cm]{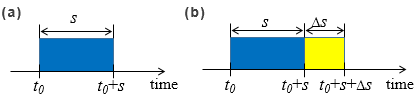}
\caption{(a) Schematic of $p_{i}(t_{0},s)$ and (b) Schematic of $q_{i}(t_{0},s+\Delta s)$}
\label{fig:C.1}
\end{figure}

Let's denote:
\begin{itemize}
\item $p_{i}(t_{0}, s)$: Probability of individual $i$ being infected from entering the restroom at time $t_0$ until time $t_{0}+s$ (refer to Figure \ref{fig:C.1}).
\item $q_{i}(t_{0}, s)$: Probability of individual $i$ remaining uninfected during the time from entering the restroom at $t_0$ until $t_{0}+s$.
\item $\beta_{i}(t_{0}+s)\Delta s$: Probability of  individual $i$ being infected within the time interval $[t_{0}+s, \Delta s]$.
\end{itemize}

By definition, $q_{i}(t_{0}, s+\Delta s)$ represents the probability of individual $i$ remaining uninfected after a duration $s+\Delta s$ from $t_0$. It can be segmented into two parts using the multiplication rule of probability: $q_{i}(t_{0},s)$, the probability that individual $i$ remains uninfected after a duration $s$, multiplied by $q_{i}(t_{0}+s,\Delta s)$, the probability that individual $i$ remains uninfected in the subsequent $\Delta s$. This process is illustrated in Figure \ref{fig:C.1}. Thus
\begin{equation}
\label{I_solution3}
\begin{array}{lll}
q_{i}(t_{0},s+\Delta s) &=& q_{i}(t_{0},s)q_{i}(t_{0}+s,\Delta s)\\
&=& q_{i}(t_{0},s)(1-\beta_{i}(t_{0}+s)\Delta s) \\
&=& q_{i}(t_{0},s)-\beta_{i}(t_{0}+s)q_{i}(t_{0},s)\Delta s
\end{array}
\end{equation}
Organizing the above equation, we derive:
\begin{eqnarray*}
q_{i}(t_{0},s+\Delta s)- q_{i}(t_{0},s)&=& -\beta_{i}(t_{0}+s)q_{i}(t_{0},s)\Delta s\\
\lim_{\Delta s\to 0} \dfrac{q_{i}(t_{0},s+\Delta s)- q_{i}(t_{0},s)}{\Delta s}&=&-\beta_{i}(t_{0}+s)q_{i}(t_{0},s)\\
\dfrac{\partial q_{i}(t_{0},s)}{\partial s}&=&-\beta_{i}(t_{0}+s)q_{i}(t_{0},s).
\end{eqnarray*}
Given that $q_{i}(t_{0},0)=1$ (indicating individual $i$ is uninfected upon entering the restroom), we obtain the differential equation
$$\dfrac{\partial q_{i}(t_{0},s)}{\partial s}=-\beta_{i}(t_{0}+s)q_{i}(t_{0},s),\quad q_i(t_0, 0) = 1.$$
Solving this differential equation (with $t_0$ as a parameter), we find:
$$q_{i}(t_{0},s)=\exp\left[{-\int_{0}^{s}\beta_{i}(t_{0}+s')ds'}\right]. $$
Hence, the probability of individual $i$ being infected within $s$ after entering the restroom is:
$$p_{i}(t_{0},s)=1-q_{i}(t_{0},s)=1-\exp\left[{-\int_{0}^{s}\beta_{i}(t_{0}+s')ds'}\right].$$
In other words:
\begin{equation}
\label{eq:30}
p_{i}(t_0, s)=1-\exp\left[{-\int_{t_{0}}^{t_{0}+s}\beta_{i}(s')ds'}\right].
\end{equation}

Using the previously calculated $\beta_i(t)$:
\begin{equation}
\label{eq:26}
  \beta_{i}(t)=\beta_{\max(i)}(t) M_i (1-V_{i}),
\end{equation}
we have
$$\beta_i(t_0 +s) = \beta_{\max(i)}(t_0+s) M_i (1-V_i),$$
where $\beta_{\max(i)}(t_0 + s)$ is defined by:
\begin{equation}
\label{eq:27}
  \beta_{\max(i)}(t)=\beta_{\max}\frac{(EV(t)/E_{0})^{4}}{1+(EV/E_{0})^{4}}.
\end{equation}

Additionally, we must consider situations where susceptible individuals are infected by objects in the restroom. Assuming a susceptible individual enters the restroom at $t_0$ and stays for a duration $s$, the time interval during which they are in the restroom is $(t_0, t_0 + s)$. Using the comprehensive infection probability $\mu_i(t_0, s)$ derived previously, the probability of susceptible individual $i$ remaining uninfected can be calculated as
$$1 - \mu_i(t_0, s).$$
Thus, the probability of susceptible individual $i$ remaining uninfected in the restroom after entering at $t_0$ and staying for $s$ is:
$$q_i(t_0, s) (1 - \mu_i(t_0, s).$$
Consequently, the infection rate of individual $i$ is:
\begin{equation}
\label{eq:31}
p_{i}(t_0, s) = 1-\exp\left[{-\int_{t_{0}}^{t_{0}+s}\beta_{i}(s')ds'}\right] \times (1 - \mu_i(t_0, s)).
\end{equation}

Ultimately, if an individual contaminates objects within the restroom or becomes infected by contaminated objects after entering, relying solely on a formula to compute the infection rate might lead to logical confusion. Hence, we recommend simulating this process based on each individual's entry into the restroom rather than relying solely on formulaic calculations. By simulating the process, the actual infection rate can be accurately determined. The steps and implementation details for simulating this process can be elucidated based on the formula derivation process outlined above.

\vspace{10pt} \noindent
{\bf Acknowledgements:}
The authors thank the helpful discussions with Dr. Xiaopeng Qi from the Chinese Center for Disease Control and Prevention. This work was supported by the National Natural Science Foundation of China under grant No.11831015.


\begin{thebibliography}{10}
\expandafter\ifx\csname url\endcsname\relax
  \def\url#1{\texttt{#1}}\fi
\expandafter\ifx\csname urlprefix\endcsname\relax\def\urlprefix{URL }\fi
\expandafter\ifx\csname href\endcsname\relax
  \def\href#1#2{#2} \def\path#1{#1}\fi

\bibitem{WorldHealthOrganization2020a}
{World Health Organization}, {Press Conference of WHO-China Joint Mission on
  COVID-19} (2020).

\bibitem{WorldHealthOrganization2020b}
{World Health Organization}, {Report of the WHO-China Joint Mission on
  Coronavirus Disease 2019} (2020).

\bibitem{2020Epidemic}
Y.~Huang, L.~Yang, H.~Dai, F.~Tian, K.~Chen, {Epidemic situation and
  forecasting of COVID-19 in and outside China} (2020).

\bibitem{2020Estimation}
B.~Tang, X.~Wang, Q.~Li, N.~L. Bragazzi, S.~Y. Tang, Y.~N. Xiao, J.~H. Wu,
  {Estimation of the transmission risk of 2019-nCov and its implication for
  public health interventions}, J. Clin. Med. 9 (2020) 462.

\bibitem{2020A}
Y.~Chen, J.~Cheng, Y.~Jiang, K.~Liu, {A time delay dynamical model for outbreak
  of 2019-nCOV and the parameter identification}, Journal of Inverse and
  III-posed Problem 28 (2020) 243--250.

\bibitem{2020Modeling}
Y.~Yue, C.~Yu, L.~Keji, {Modeling and prediction for the trend of outbreak of
  NCP based on a time-delay dynamic system}, Scientia Sinica Mathematica 50~(3)
  (2020) 385.

\bibitem{Deng:2021gg}
Y.~Deng, S.~Xing, M.~Zhu, J.~Lei, {Impact of insufficient detection in COVID-19
  outbreaks}, Math Biosci Eng 18~(6) (2021) 9727--9742.

\bibitem{2021Effectiveness}
C.~Xu, Y.~Pei, X.~Qi, S.~Liu, J.~Lei, Effectiveness of non-pharmaceutical
  intervention against local transmission of covid-19: An individual-based
  modelling study, SSRN Electronic Journal 6 (2021) 848--858.

\bibitem{2020DataDriven}
T.~Alamo, D.~G. Reina, P.~M. Gata, V.~M. Preciado, G.~Giodano, {Data-diven
  methods for presnet and future pandemics: Monitoring, modelling and
  managing}, Annu Rev Control 52 (2021) 448--464.

\bibitem{2017Using}
S.~Venkatramanan, B.~Lewis, J.~Chen, D.~Higdon, A.~Vullikanti, M.~Marathe,
  Using data-driven agent-based models for forecasting emerging infectious
  diseases, Epidemics 22 (2018) 43--49.

\bibitem{2021From}
D.~Bertsimas, L.~Boussioux, R.~Cory-Wright, A.~Delarue, V.~Digalakis,
  A.~Jacquillat, D.~L. Kitane, G.~Lukin, M.~Li, L.~Mingardi, From predictions
  to prescriptions: A data-driven response to covid-19, Health Care Management
  Science 2 (2021) 1--20.

\bibitem{2020Data}
B.~Tang, S. Y.~Tang, N.~L. Bragazzi, Data mining of covid-19 epidemic and
  analysis of discrete random propagation dynamic model, Science china 50
  (2020) 1--16.

\bibitem{Ajelli2010ComparingLC}
M.~Ajelli, B.~Gonalves, D.~Balcan, V.~Colizza, A.~Vespignani, Comparing
  large-scale computational approaches to epidemic modeling: Agent-based versus
  structured metapopulation models, BMC Infectious Diseases 10~(1) (2010)
  1--13.

\bibitem{Xu2021EffectivenessON}
C.~Xu, Y.~Pei, S.~Liu, J.~Lei, Effectiveness of non-pharmaceutical
  interventions against local transmission of covid-19: An individual-based
  modelling study, Infectious Disease Modelling 6 (2021) 848 -- 858.

\bibitem{Chang2020ModellingTA}
S.~L. Chang, N.~Harding, C.~Zachreson, O.~M. Cliff, M.~Prokopenko, Modelling
  transmission and control of the covid-19 pandemic in australia, Nature
  Communications 11 (2020).

\bibitem{Wu:2023kg}
F.~Wu, X.~Liang, J.~Lei, {Modelling COVID-19 epidemic with confirmed
  cases-driven contact tracing quarantine}, Infectious Disease Modelling 8
  (2023) 415--426.

\end{thebibliography}

\end{document}